\documentclass{article}

\usepackage{lineno}
\usepackage{authblk}
\usepackage{amsmath}
\usepackage{amssymb}
\usepackage{graphicx}
\usepackage{caption}
\usepackage{subcaption}
\usepackage{bm}
\usepackage{booktabs}
\usepackage[hidelinks]{hyperref}
\usepackage[round, authoryear]{natbib}
\usepackage{chapterbib}
\usepackage[margin=2.5cm]{geometry}
\usepackage{setspace}
\usepackage{booktabs}
\usepackage{etoolbox} 

\newcommand*\linenomathpatch[1]{%
  \cspreto{#1}{\linenomath}%
  \cspreto{#1*}{\linenomath}%
  \csappto{end#1}{\endlinenomath}%
  \csappto{end#1*}{\endlinenomath}%
}

\linenomathpatch{equation}
\linenomathpatch{gather}
\linenomathpatch{align}

\graphicspath{{./graphics/}}

\newcommand{\calD}{\mathcal D}

\newcommand{\bA}{ {\boldsymbol A} }
\newcommand{\bb}{ {\boldsymbol b} }

\newcommand{\bD}{ {\boldsymbol D} }

\newcommand{\bH}{ {\boldsymbol H} }

\newcommand{\bI}{ {\boldsymbol I} }

\newcommand{\bp}{ {\boldsymbol p} }

\newcommand{\bQ}{ {\boldsymbol Q} }

\newcommand{\bs}{ {\boldsymbol s} }

\newcommand{\bw}{ {\boldsymbol w} }

\newcommand{\bx}{ {\boldsymbol x} }
\newcommand{\bX}{ {\boldsymbol X} }
\newcommand{\by}{ {\boldsymbol y} }

\newcommand{\bz}{ {\boldsymbol z} }

\newcommand{\bone}{ {\bf 1} }
\newcommand{\bzero}{ {\bf 0} }

\newcommand{\bbeta}{ {\boldsymbol \beta} }

\newcommand{\bepsilon}{ {\boldsymbol \epsilon} }

\newcommand{\bpi}{ {\boldsymbol \pi} }

\newcommand{\bet}{ {\boldsymbol \eta} }

\newcommand{\bSigma}{ {\boldsymbol \Sigma} }

\newcommand{\bpsi}{ {\boldsymbol \psi} }

\newcommand{\iid}{\overset{\mbox{iid}} \sim}

\begin{document}
\begin{center}
        \vspace*{1cm}
        \large
	    \textbf{A spatial mixture model for spaceborne lidar observations over mixed forest and non-forest land types}\\
         \normalsize
        \vspace{5mm}
	    Paul B. May\textsuperscript{1,2}, Andrew O. Finley\textsuperscript{3,4}, and Ralph O. Dubayah\textsuperscript{1}\\
    
         \vspace{5mm}
        $^1$Department of Geographical Sciences, University of Maryland, College Park, MD, USA.\\
        $^2$Department of Mathematics, South Dakota School of Mines \& Technology, Rapid City, SD, USA. \\
        $^3$Department of Forestry, Michigan State University, East Lansing, MI, USA. \\
        $^4$Department of Statistics and Probability, Michigan State University, East Lansing, MI, USA.\\

 \end{center}
\vspace*{2mm}

\begin{abstract}
The Global Ecosystem Dynamics Investigation (GEDI) is a spaceborne lidar instrument that collects near-global measurements of forest structure. While expansive in scope, GEDI samples are spatially sparse and cover a small fraction of the land surface. Converting the sparse samples into spatially complete predictive maps is of practical importance for a number of ecological studies. A complicating factor is that GEDI collects measurements over forested and non-forested land alike, with no automatic labeling of the land type. Such classification is important, as it categorically influences the probability distribution of the spatial process and the ecological interpretation of the observations/predictions. We propose and implement a spatial mixture model, separating the observations and the greater spatial domain into two latent classes. The latent classes are governed by a Bernoulli spatial process, with spatial effects driven by a Gaussian process. Within each class, the process is governed by a separate spatial model, describing the unique probabilistic attributes. Model predictions take the form of scalar predictions of the GEDI observables as well as discrete labeling of the class membership. Inference is conducted through a Bayesian paradigm, yielding rich quantification of prediction and uncertainty through posterior predictive distributions. We demonstrate the method using GEDI data over Wollemi National Park, Australia, using optical data from Landsat 8 as model covariates. When compared to a single spatial model, the mixture model achieves much higher posterior predictive densities on the true value. When compared to a random forest model, a common algorithmic approach in the remote sensing community, the random forest achieves better absolute prediction accuracy for prediction locations far from observed training data locations, but at the expense of location-specific assessments of uncertainty. The unsupervised binary classifications of the mixture model appear broadly ecologically interpretable as forest and non-forest when compared to optical imagery, but further comparison to ground-truth data is required.
\end{abstract}

\section{Introduction}

Lidar (light detection and ranging) has become an indispensable tool in forest ecosystem studies \citep{dubayah2000lidar}. Lidar instruments are lasers that emit pulses of light that reflect from an impacted object, measuring the time to return and the shape of returned waveform to gauge the position and structure of the impacted object. The use of lidar instruments and other remote sensing technologies in conjunction with \textit{in situ} field measurements has sparked a revolution in forestry and environmental sciences by allowing estimates of important forest attributes at finer resolutions and broader spatial scales than was possible with the previous paradigm of relying on field measurements alone \citep{akay2009using}.\par 
Airborne laser scanning (ALS) is a common lidar implementation comprised of a lidar instrument mounted on an aircraft, flown over areas of interest to measure vegetation and topography. While powerful in their ability to provide spatially continuous measurements of forest structure, individual ALS campaigns are often limited in scope, covering small geographic areas \citep{wulder2012lidar}, motivating spaceborne lidar sampling missions such as the Global Ecosystem Dynamics Investigation (GEDI).\par 
GEDI is a lidar instrument onboard the International Space Station providing estimates of forest structure between latitudes $51.6^\circ$ N and $51.6^\circ$ S \citep{dubayah2020global}. While GEDI and other spaceborne lidar instruments can provide observations that are expansive in scope, the observations are not spatially continuous, or `wall-to-wall'. Instead, observations are discretely sampled `footprints' that cover a small fraction of the Earth's surface. Discrete laser pulses emitted from GEDI illuminate footprints on Earth's surface approximately 25 m in diameter. A reflected waveform is returned to GEDI, from which the vertical arrangement of matter can be inferred. When the laser illuminates a forested footprint, the returned waveform is indicative of the vertical arrangement of plant matter. For example, the RH98 metric (98\% percentile relative height) gives the height above the estimated ground below which 98\% of the waveform energy was reflected. For forested footprints, RH98 provides a proxy for forest canopy height. However, in general there is no way to know \textit{a priori} whether the footprint is forested, and steep slopes/rough topography in non-forested areas can produce large RH98 values that could be confused for canopy heights, producing misleading estimates of forest biomass or habitat suitability \citep[Section 4]{dubayah2022gedi}.\par
The primary objective of the GEDI mission is to use the discrete samples of GEDI to provide design-based estimates of forest attribute population means at a 1 km resolution. However, there is great interest in leveraging GEDI data to produce spatially complete predictive maps at finer resolutions, which are then used for tree and animal species modeling \citep{marselis2019exploring, burns2020incorporating, marselis2022use}, characterizing forest fuels for wildfire management \citep{hoffren2023assessing}, or estimating forest biomass \citep{qi2019forest} at fine scales. The dominant method for producing these predictive maps is fusion with auxiliary remote sensing sources that are spatially complete (typically passive optical or radar data) through regression models \citep{qi2016combining, potapov2021mapping, sothe2022spatially}. Models are trained where the GEDI observations and auxiliary sources overlap, and then the spatial continuity of the auxiliary data is used to make spatially complete predictions. These methods are typically algorithmic and do not provide location-specific quantification of uncertainty, but rather give cross-validation statistics as a general measure of accuracy. Rigorous quantification of prediction uncertainty is not only crucial for interpretation of the predicted metrics themselves, but for tracking and accounting for uncertainties in downstream ecological models that utilize the predicted metrics \citep{calder2003incorporating, barry2006error, molto2013error}.\par
The general problem statement, leveraging spatially incomplete observations to make spatially complete predictions, is far from foreign in the spatial statistics community. The typical model for such objectives is a (generalized) linear spatial model, where the mean of the observed process is given by a linear regression on a set of covariates plus a Gaussian process (GP) spatial error, which describes spatially patterned variation not explained by the covariates \citep{gelfand2016spatial}. Predictions at unobserved locations utilize the covariates as well as the spatial covariance of the Gaussian process. Within the Bayesian paradigm, predictions take the form of posterior distributions, providing scalar predictions (e.g., posterior expected values and modes) and rich quantification of uncertainty.\par 
A common assumption is that the spatial process is stationary conditional on the covariates, where the regression coefficients are constant over the study domain and the covariance of the GP is invariant to translation and only depends on the relative distances between locations. However, many ecosystems are heterogeneous mixtures of forested and non-forested areas, and we expect categorical differences in the stochastic behavior of observed GEDI metrics between these land types. Assumptions of stationarity have been often relaxed, with spatially varying regression coefficients \citep{gelfand2003spatial} and non-stationary covariance functions for the GP error \citep{sampson1992nonparametric, paciorek2006spatial, bolin2011spatial}. These methods assume smooth spatial variation in the model parameters or continuous transformations of the coordinate space, but we hypothesize discrete breaks in the behavior between the forest/non-forest classes and homogeneous behavior within, i.e. a mixture of two stationary spatial processes. \cite{finley2011hierarchical} employed a hierarchical model for predicting continuous forest variables over mixed forest/non-forest areas, but observations were \textit{in situ} field plot data with the classification labels known at the observation locations.\par
Spatial mixture models with latent class membership have been developed in the literature. \cite{wall2009spatial} clustered spatial multivariate binary data, using GPs to account for spatial dependence within the class probabilities and assuming observations were independent given their class membership. This was extended to clustering spatio-temporal multivariate binary observations \citep{vanhatalo2021spatiotemporal}, again using GPs to drive class probabilities and assuming independent observations given the class membership. \cite{neelon2014multivariate} developed a spatial mixture model for continuously-valued areal data, where the class probabilities were spatially correlated through a conditional autoregressive (CAR) process, and the within-class observations were correlated through an additional CAR process.\par
We extend these developments to model and predict point-referenced (log)-RH98 from GEDI across a heterogeneous landscape in Wollemi National Park, Australia. A latent Bernoulli linear spatial model is used to model hypothesized forest/non-forest classifications, mixing two separate linear spatial models for RH98. We use passive optical data from the Landsat 8 satellite as model covariates, influencing both class membership and RH98 within the classes. In order to accommodate the magnitude of the GEDI data, the stochastic partial differential equation (SPDE) approach is employed to model the GP, where the GP is represented by the projection of a fixed-rank Gaussian Markov random field with a sparse precision matrix. Inference is conducted through a Bayesian paradigm, using a Gibbs sampler for model posteriors and a Laplace approximation for the conditional samples of the linear effects within the Bernoulli model.\par 

The posited spatial mixture model acknowledges the different properties of the classes, which has a significant impact on the predictive inference of continuously-valued RH98, but also provides categorical prediction of the class membership, which should influence how the values are interpreted. For instance, predicted or observed RH98 over non-forested areas, indicative of topography but not canopy height, would not be interpreted as positive predictors of forest biomass or suitable tree-dwelling bird habitats. We compare the spatial mixture model's predictive inference to a single linear spatial model, showing the mixture model on average provides much higher posterior density at the true value. The spatial mixture model provides visually clear separation of forest and non-forest areas when compared to optical imagery, but further investigation is required to deduce the accordance of the unsupervised classification to ground-truth definitions of forested land.           

\section{Data and Study Area}

\begin{figure}[htb]
	\centering
	\begin{subfigure}[b]{0.48\textwidth}
		\centering
		\includegraphics[width = \textwidth, keepaspectratio]{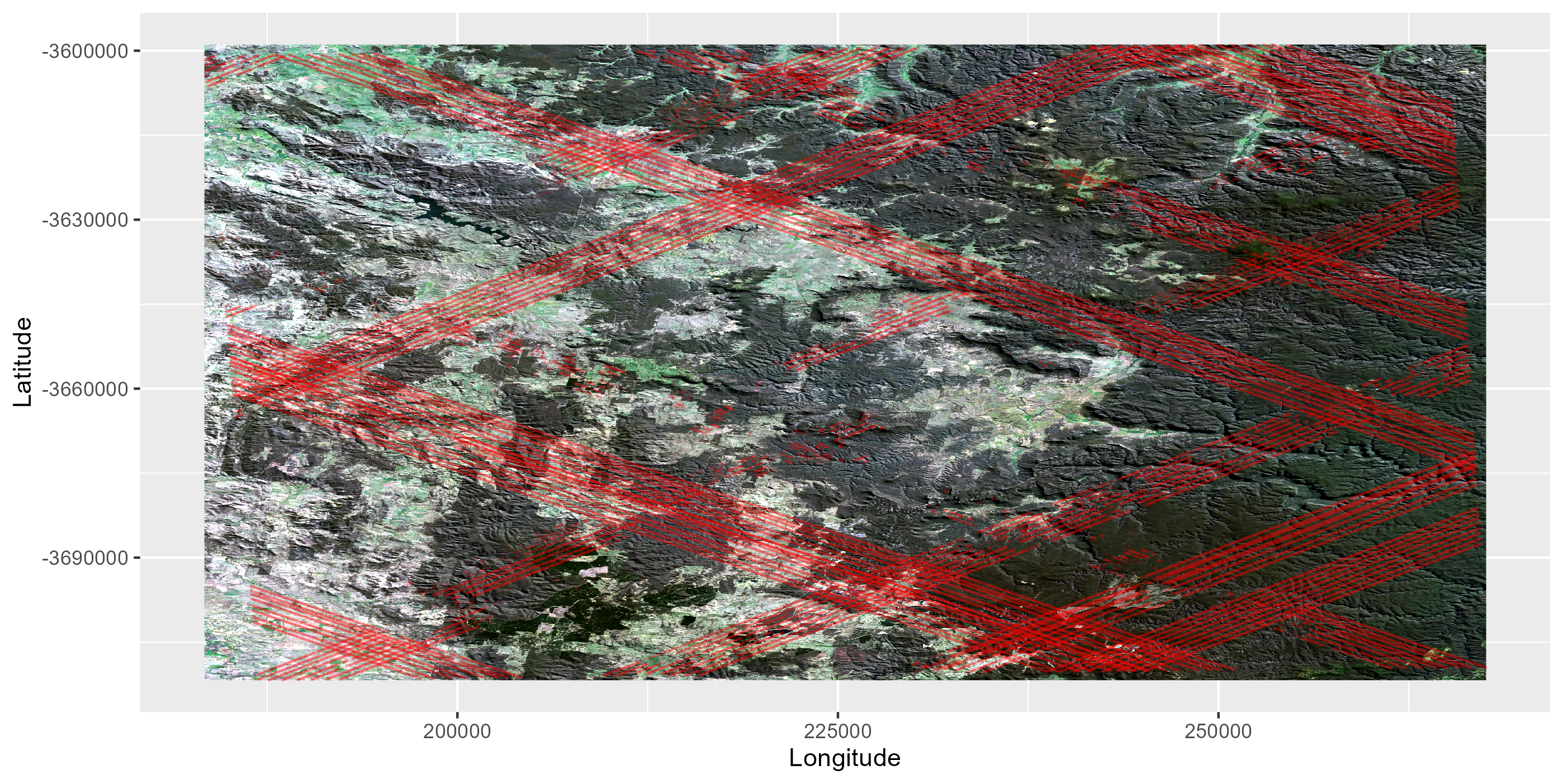}
        \caption{}
        \label{fig:datadescripA}
	\end{subfigure}
	\hfill
	\begin{subfigure}[b]{0.48\textwidth}
		\centering
		\includegraphics[width = \textwidth, keepaspectratio]{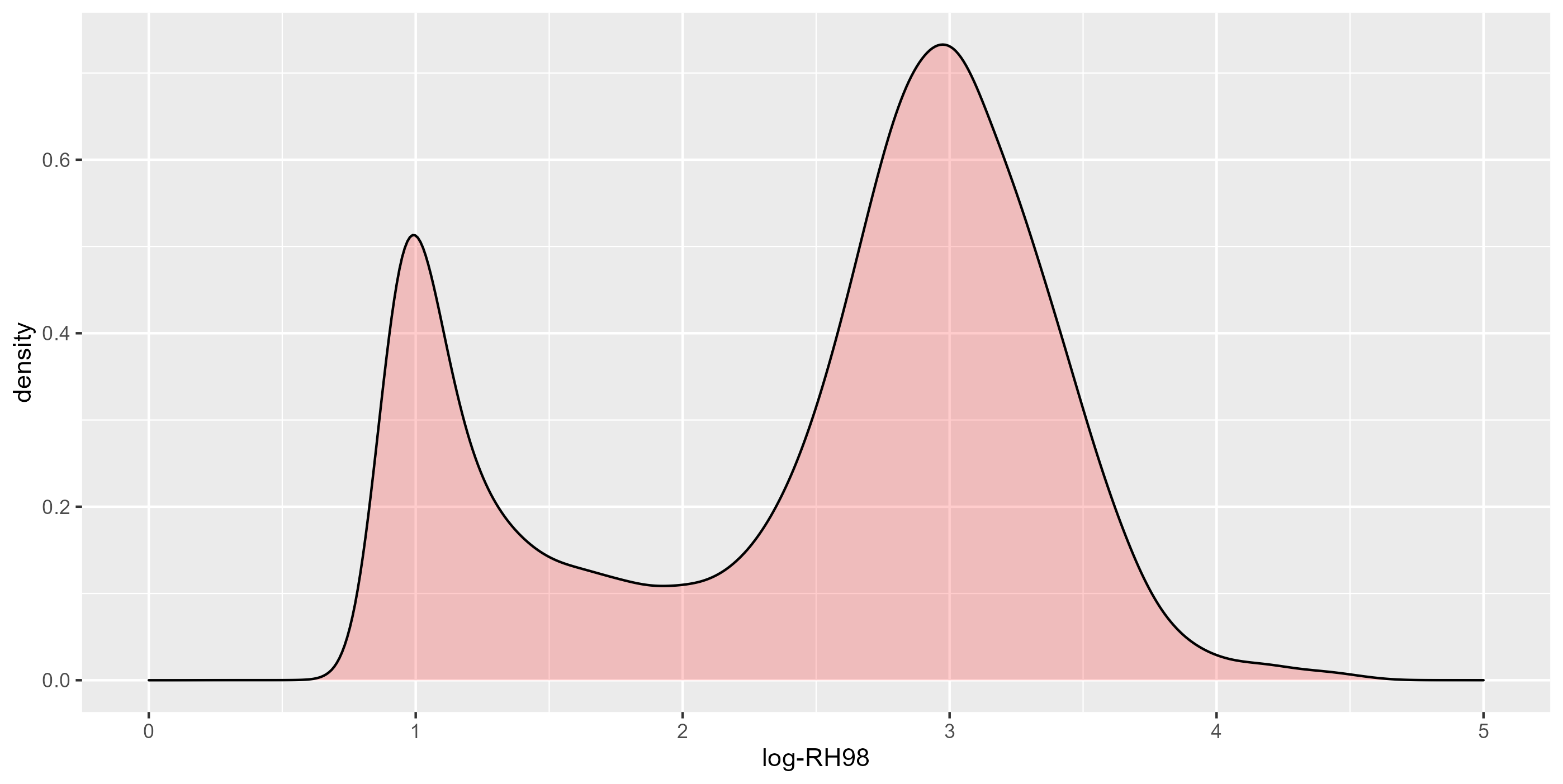}
        \caption{}
        \label{fig:datadescripB}
	\end{subfigure}
	\caption{The  85 x 115 km region in Wollemi National Park, Australia: (a) A RGB (red-green-blue) rendering of the Landsat image with the observed GEDI footprints plotted over-top in red. (b) The empirical distribution of the observed log-RH98, produced with a kernel density estimate using a Gaussian kernel and a bandwidth $<0.1$. The GEDI footprints cover both forested and non-forested areas, which is strongly manifested in the bimodal empirical distribution.}\label{fig:datadescrip}
\end{figure}

Our study area is a 85 x 115 km region in Wollemi National Park, Australia (Figure~\ref{fig:datadescripA}). This study area is of particular interest due to the unprecedented wildfires that occurred during 2019--2020 \citep{smith2021quantifying}. Baseline estimates of forest attributes pre-fire are critical for accurate estimation of ecological impact, but impossible to collect in retrospect, making historical satellite data a useful tool. GEDI collects lidar observations from 25 m diameter footprints along 8 parallel ground tracks following its orbit path, with 60 m between footprints along-track, and 600 m between tracks. From the beginning of its mission in March 2019, GEDI collected 94,513 quality footprint observations over the study area before the fires. To better meet normality assumptions and maintain positive support (after back-transformation), we model the logarithm of the RH98 metric as the response variable. Inference on the original scale can easily be reclaimed by back-transforming posterior predictive samples, though some biomass models, for instance, do use the log-RH98 values \citep{duncanson2022aboveground}. For covariates, we use pre-fire imagery from Landsat 8, which gives optical intensities in $p = 7$ different bands (ranges of light wavelengths) in a spatially complete grid with a 30 x 30 m resolution. The study area is a mixture of forested and non-forested areas, which is visually evident from the Landsat image and strongly manifested in the bimodal empirical distribution of the log-RH98 values (Figure \ref{fig:datadescrip}). We hypothesize the right distribution is largely representative of forest observations, while the left distribution is largely representative of non-forest observations.

\section{Methods}\label{sec:methods}

\subsection{Typical spatial model}

Let $y(s)$ be the log-RH98 at location $s\in\calD$, which is observed incompletely across study domain $\calD\subset \mathbb{R}^2$, and $\bx(s)$ be the $p\times 1$ vector of collocated optical intensities, observed completely across $\calD$. The objective is inference on $y(s)$ for all $s\in \calD$. A ubiquitous model for such data and objectives is
\begin{equation}\label{eq:typical}
y(s) = \mu + \bx(s)^T\bbeta + \eta(s) + \epsilon(s),
\end{equation}
where $\mu$ is a constant scalar intercept, $\bbeta$ is a $p\times 1$ vector of regression coefficients, $\eta(s)$ is a mean-zero Gaussian process (GP), and $\epsilon(s)$ is spatially independent Gaussian noise with mean zero and constant variance $\tau^2$. We assign a Mat\'ern covariance function to the GP \citep[p. 48]{stein1999interpolation}, with fixed smoothness $\nu = 1$ and variance and range parameters $\theta = \{\sigma^2,\phi\}$ respectively:
\begin{equation}\label{eq:matern}
\mathrm{Cov}[\eta(s), \eta(s')] = C_{\nu=1}(\|s - s'\|~;~\theta) = \sigma^2\sqrt{8}\frac{\|s-s'\|}{\phi}K_1\left(\sqrt{8}\frac{\|s-s'\|}{\phi}\right)
\end{equation}
where $K_1(\cdot)$ is an order $1$ modified Bessel function of the second kind and $\|\cdot\|$ is the Euclidean distance. We use the parameterization of \citep{lindgren2011explicit} in (\ref{eq:matern}), as range parameter $\phi$ is then interpretable as the distance at which the correlation is approximately 0.13.\par 
 Due to the size of the data, we approximate all GPs with the stochastic partial differential equation (SPDE) approach \citep{lindgren2011explicit}; more on this in Section \ref{sec:SPDE}. We refer to (\ref{eq:typical}) as the typical spatial model. Inference on $y(s_*)$ at some unobserved location $s_*$ leverages optical intensities $\bx(s_*)$ as well as the posterior of $\eta(s_*)$, which is informed by nearby observations.

\subsection{Spatial mixture model}

The typical spatial model in (\ref{eq:typical}) posits a single spatial process with a linear relationship with the optical intensities, augmented by spatial and non-spatial errors, $\eta(s),~\epsilon(s)$, with constant variance. In addition to a constant variance, the Gaussian process $\eta(s)$ has a single range parameter, so that posteriors are influenced only by the distances relative to the observations and the scalar magnitude of those observations. We can imagine some shortcomings of this model for the data. The study area, like many ecosystems, is a mixture of forested and non-forested areas. These classes are likely to have widely different statistical properties. For one, rather than a single linear relationship with the optical intensities $\bx(s)$ that spans both classes, it may be advantageous to pose a discrete break in the relationship, with different regression coefficients for each class. Further, the structure of the errors, described by the variances and the range of the GP and the variance of the noise, may be quite different. It is also reasonable to suppose breaks in the spatial correlation of the GP, where two observations from different classes are spatially independent, even if immediately adjacent. The challenge is that delineations between classes are not known \textit{a priori} at GEDI observed locations or at desired prediction locations. Thus the class membership is latent, and the model must separate the processes through `unsupervised' classification.\par
Let $z(s)$ be a binary variable giving the class at location $s$, where $z(s) = 1$ indicates our hypothesized forest class and $z(s) = 0$ otherwise. The spatial mixture model proposed in this work is given by
\begin{align}
z(s)|\pi(s) &\sim \mathrm{Bernoulli}\left(\pi(s)\right),\label{eq:fprocess}\\ 
\mathrm{logit}\left(\pi(s)\right) &= \mu_z + \bx(s)^T\bbeta_z + \eta_z(s),\label{eq:pprocess}\\
y_1(s) \equiv y(s)|\left(z(s)=1\right) &= \mu_1 + \bx(s)^T\bbeta_1 + \eta_1(s) + \epsilon_1(s)\label{eq:1process},\\
y_0(s) \equiv y(s)|\left(z(s)=0\right) &= \mu_0 + \bx(s)^T\bbeta_0 + \eta_0(s) + \epsilon_0(s)\label{eq:0process},
\end{align}
giving a mixture of two spatial processes, $y_1(s),~y_0(s)$, the mixing dictated by Bernoulli $z(s)$, for which the probability is given by the spatial process $\pi(s)$. Parameters $\mu_z,\mu_1,\mu_0$ and $\bbeta_z,\bbeta_1,\bbeta_0$ are unique intercepts and regression coefficients, respectively. Effects $\eta_z(s),\eta_1(s), \eta_0(s)$ are GPs, independent from each other, with Mat\'ern covariance (\ref{eq:matern}) and parameters $\theta_z = \{\sigma_z^2,\phi_z\}$, $\theta_1 =\{\sigma_1^2,\phi_1\}$, $\theta_0 = \{\sigma_0^2,\phi_0\}$, and $\epsilon_1$ and $\epsilon_0$ are spatially independent Gaussian noise with mean zero and variances $\tau_1^2$ and $\tau_0^2$, respectively.\par
In the proposed model, processes $\pi(s)$, $y_1(s)$ and $y_0(s)$ are independent from each other. It is possible to change this with the introduction of shared GP effects, e.g.,
\begin{equation}
y_1(s) = \mu_1 + \bx(s)^T\bbeta_1 + \eta_1(s) + \alpha\eta_z(s) + \epsilon_1(s),
\end{equation}
which would induce covariance between $\mathrm{logit}\left(\pi(s)\right)$ and $y_1(s)$, depending on the value of scalar $\alpha$. When tested, we saw no obvious improvements from such inclusions for our data.

\subsection{SPDE representation of the Gaussian process}\label{sec:SPDE}

A primary obstacle to modeling with GPs is the computational burden associated with the covariance matrix of observations. Given $n$ locations $s_1,\ldots,s_n$, the vector $[\eta(s_1), \ldots, \eta(s_n)]^T$ has a multivariate normal prior distribution with an $n\times n$ covariance matrix. For a generic covariance function, this covariance matrix will be dense, requiring operations of $O(n^3)$ computational complexity and $O(n^2)$ memory usage to evaluate the prior or marginal likelihood. Such would be infeasible with our data, with almost 100,000 observations. Substantial work has been dedicated to making GPs computationally accessible for large data sets, consisting of various approximations and carefully constructed covariance functions \citep{heaton2019case}. In this work, we use the SPDE approach to model the GPs \citep{lindgren2011explicit}.\par
For a detailed development of the SPDE approach, the reader is pointed to \cite{lindgren2011explicit}. We provide only the general overview required to understand its implementation in this work. The SPDE approach is founded on the fact that a GP with Mat\'ern covariance is the stationary solution to a SPDE. The solution can be represented using the finite element method, where the finite elements are constructed as a Delaunay triangulation, or mesh, over the study area, partitioning the area into non-overlapping triangles. A Gaussian Markov random field $w_j;~j\in\{1,\ldots,k\}$ is defined on $k$ triangle vertices, with a $k\times k$ precision matrix $\bQ(\theta)$ that is a function of the Mat\'ern parameters, $\theta$, and geometry of the mesh. Matrix $\bQ(\theta)$ is extremely sparse, and the non-zero entries can be computed in closed-form. The continuously-indexed GP $\eta(s)$ is then defined by
\begin{equation}
\eta(s) = \sum_{j=1}^k a_j(s)w_j,
\end{equation}
for basis functions $a_j(s);~j\in \{ 1,\ldots, k \}$. The basis functions are piece-wise linear functions defined by the mesh, where location $s$ contained in the triangle with vertices $j\in\{1,2,3\}$, for example, will have non-zero coefficients $a_1(s), a_2(s), a_3(s)$ summing to one, while the remaining $a_j(s)$ are zero. If $s$ lies on a triangle edge connecting vertices $j\in \{1, 2\}$, then $a_1(s), a_2(s)$ are non-zero, summing to one, with the remaining $a_j(s)$ zero, while if $s$ is on vertex $j = 1$, then $a_1(s)=1$ and the remaining $a_j(s)$ are zero. Therefore, for all $s$, we have $\sum_{j=1}^k a_j(s) = 1$, with at most three non-zero entries. For a vector of $n$ locations $\bs = [s_1 \cdots s_n]^T$, we have
\begin{equation}
\bet(\bs)= \begin{bmatrix} \eta(s_1) \\ \vdots \\ \eta(s_n) \end{bmatrix} = \bA(\bs)\bw; \text{ where } \bw = \begin{bmatrix} w_1 \\ \vdots \\ w_k \end{bmatrix},
\end{equation}
and where $\bA(\bs)$ is a sparse projection matrix such that $[\bA(\bs)]_{ij} = a_j(s_i)$. Thus any multivariate realization of $\eta(s)$ is a projection of fixed-dimension $\bw$.\par 
For our analysis, we use the same mesh for all GPs: $\eta(s)$ in the typical model and $\eta_z(s), \eta_1(s), \eta_0(s)$ in the mixture model. For vector of locations $\bs$, we have
\begin{equation}
\bet(\bs) = \bA(\bs)\bw,~~\bet_z(\bs) = \bA(\bs)\bw_z~~\bet_1(\bs) = \bA(\bs)\bw_1,~~\bet_0(\bs) = \bA(\bs)\bw_0
\end{equation}
where the properties of $\bw,\bw_z,\bw_1,\bw_0$ are distinguished by their respective Mat\'ern parameters, $\theta,\theta_z,\theta_1,\theta_0$. We impose a maximum triangle edge length of 1,000 m, a minimum edge length of 500 m, and buffer with a 7,000 m radius, so that the mesh extends beyond the boundary of observations by at least 7,000 m in any direction. This results in a mesh with $k = 25,276$ vertices. We generated the mesh and computed the associated projection matrices $\bA(\cdot)$ and precision matrices $\bQ(\theta)$ using `R' package `INLA' \citep{rcore, rinla}. 

\subsection{Bayesian model inference and prediction}

We conduct Bayesian inference on both the typical and mixture model using Gibbs sampling. In this section, we outline the inference procedure for both models and the procedure for producing posterior predictive distributions for $y(s)$ at unobserved locations. Exact details on the Gibbs sampler can be found in Appendix \ref{app:gibbs}. 

\subsubsection{Inference for the typical spatial model}\label{sec:typicalgibbs}

The unknown quantities to be sampled are parameters $\mu,\bbeta,\theta, \tau^2$ and effect $\bw$. We assign flat priors to all parameters except the Mat\'ern parameters, $\theta = \{\sigma^2, \phi\}$, which are not mutually consistently estimable for a fixed domain, $\calD$ \citep{zhang2004inconsistent}: parameter pairs with identical ratios, $\sigma^2/\phi$, will produce very similar likelihoods, and therefore informative priors are required to constrain the likelihood and produce a well-behaved posterior distribution. For $\theta$ we employ penalized complexity (PC) priors, which exert downward pressure on the variance, $\sigma^2$, and upward pressure on the range, $\phi$ \citep{fuglstad2019constructing}. Specifically, we impose prior probabilities $\mathrm{Prob}(\sigma > 1) = 0.01$ and $\mathrm{Prob}(\phi < 2,000~\text{m}) = 0.01$, where $\sigma = \sqrt{\sigma^2}$.\par 

The primary computational bottleneck in the Gibbs sampler is Cholesky decomposition of the $k \times k$ conditional precision matrix $\bQ(\theta) + \frac{1}{\tau^2}\bA(\bs)^T\bA(\bs)$, used to draw conditional samples of spatial effects $\bw$ and reused to sample the mean parameters $\mu,~\bbeta$, marginalizing over $\bw$. However, because both $\bQ(\theta)$ and $\bA(\bs)$ are exceedingly sparse, the resulting precision matrix is as well, and the decomposition can be computed relatively quickly. For reference, with our mesh, $k = 25,276$, on a 1.6 GHz laptop utilizing 2 cores, a single computation of the Cholesky decomposition takes 0.5 seconds using the sparse matrix routines in `R' package `Matrix' \citep{rmatrix}.

\subsubsection{Inference for the spatial mixture model}\label{sec:mixturegibbs}
Because the mixture model is a combination of two spatial models, we use many of the same techniques for the Gibbs sampler as with the typical spatial model. Additional complexity arises from the sampling of $\bz(\bs)$ with every iteration of the Gibbs sampler, changing which observations are attributed to each model, $y_1(s),~y_0(s)$. Further, we require inference on the parameters/effects of the probability process (\ref{eq:pprocess}) driving $\bz(\bs)$, where direct conditional samples of $\mu_z,\bbeta_z, \bw_z$ are no longer available. A Metropolis-Hastings step for the entire joint distribution would have a vanishingly small acceptance rate due to the large dimension of $\bw_z$. Alternatively, cycling through conditional samples of $w_{z,j};~j\in\{1,\ldots,k\}$ individually would induce prohibitively slow mixing due to the strong correlation between them. We instead use Laplace approximations to draw conditional samples of $[\mu_p,\bbeta_p, \bw_p]^T$, yielding approximate Bayesian model inference. For the Laplace approximation, we first implement a Newton-Raphson procedure to find the posterior mode of $\bb = [\mu_z~\bbeta_z^T~\bw_z^T]$. The Laplace approximation is then a multivariate normal distribution with mean vector given by the posterior mode and precision matrix given by the Hessian matrix at the posterior mode. \par
The unknown quantities to be sampled are parameters \[\mu_z,~\mu_1,~\mu_0,~\beta_z,~\beta_1,~\beta_0,~\theta_z,~\theta_1,~\theta_0,~\tau_1^2,~\tau_0^2\] and effects $\bz(\bs),\bw_z,\bw_1,\bw_0$. We assign flat priors to all parameters, except for the Mat\'ern parameters $\theta_z,~\theta_1,~\theta_0$, which are again given PC priors, such that
\[\begin{array}{lll}
\mathrm{Prob}(\sigma_z > 10) = 0.01, & \mathrm{Prob}(\sigma_1> 0.5) = 0.01, & \mathrm{Prob}(\sigma_0 > 0.2) = 0.01, \\
\mathrm{Prob}(\phi_z < 2,000~\text{m}) = 0.01, & \mathrm{Prob}(\phi_1 < 1,000~\text{m}) = 0.01, & \mathrm{Prob}(\phi_0 < 2,000~\text{m}) = 0.01.
\end{array}\]
Mixture models are not identifiable without informative priors, as the orientation of class labels, $z(s) \in \{0, 1\}$, is arbitrary, and could be reversed to produce an equivalent model. To generate initial values for $\bz(s)$, we used an EM algorithm, temporarily assuming a simple two-component Gaussian mixture model for $\by(\bs)$,
\begin{align}
y(s_i)|z(s_i) &\iid z(s_i)\cdot\mathrm{N}\left(\mu_{1,*},~\tau_{1,*}^2\right) + \left(1- z(s_i)\right)\cdot\mathrm{N}\left(\mu_{0,*},~\tau_{0,*}^2\right)\\
z(s_i) &\iid \mathrm{Bernoulli}(0.5)~;~~i\in\{1,\ldots,n\}, 
\end{align}
choosing initial mean parameters $\mu_{1,*} = 3,~\mu_{0,*} = 1$ for the EM algorithm to ensure the right component of the density in Figure \ref{fig:datadescrip} corresponded to class $z(s) = 1$, our hypothesized forest class. Using these initial values, we found there to be no risk for our data of the MCMC chain ``jumping" to a complete role-reversal of $z(s)$.\par
Just as with the typical model in Section \ref{sec:typicalgibbs}, the primary computational burden is Cholesky decomposition of the sparse conditional precision matrices associated with $\bw_0$, $\bw_1$ and $\bb = [\mu_z~\bbeta_z^T~\bw_z^T]$, but again these can be computed relatively quickly using sparse matrix routines. The Gibbs sampler, detailed in Appendix \ref{app:gibbs}, illustrates a particular advantage to the SPDE approach for GPs within a mixture model: even though the observations attributed to each class are potentially changed every iteration of the Gibbs sampler, the dimension of the underlying spatial effects $\bw_j;~j\in\{0,1\}$ and the form of the conditional distribution is static. This is particularly convenient for prediction, as posterior predictive predictive samples of the spatial effects $\bet_j(\bs_*)$ at a vector of locations $\bs_*$ are projections of the already-sampled $\bw_j$ using a single fixed and non-stochastic projection matrix.

\subsubsection{Prediction}

Given $M$ posterior samples of the model parameters, posterior predictive samples for $\by(\bs_*)$ at a vector of unobserved locations $\bs_*$ using the typical model can be computed as
\begin{equation}\label{eq:typpredict}
\by_{(m)}(s_*) = \mu_{(m)}\bone + \bX(\bs_*)\bbeta_{(m)} + \bA(\bs_*)\bw_{(m)} + \bepsilon_{(m)}(\bs_*);~~m\in\{1,\ldots,M\},
\end{equation}
where $\bA(\bs_*)$ is a projection matrix for the prediction locations and $\bepsilon_{(m)}(\bs_*)$ is a vector of iid normal distributed variables with mean zero and variance $\tau_{(m)}^2$. \par
For the mixture model, predictive samples of $\bpi(\bs_*),~\by_1(\bs_*),~\by_0(\bs_*)$ can be drawn in the same fashion as (\ref{eq:typpredict}). Then the predictive samples of the class membership $\bz(\bs_*)$ and continuous response $\by(\bs_*)$ are given by
\begin{align}
\bz_{(m)}(\bs_*) &\sim \mathrm{Bernoulli}\left(\bpi_{(m)}(\bs_*)\right) \\
\by_{(m)}(\bs_*) &= \bz_{(m)}(\bs_*) \odot \by_{1,(m)}(\bs_*) + \left(1 - \bz_{(m)}(\bs_*)\right)\odot\by_{0,(m)}(\bs_*),
\end{align}
for $m\in\{1,\ldots,M\}$, where $\odot$ represents element-wise multiplication.\par
Posterior predictive samples for any transformation of $\by(s_*)$ are easily achieved by applying the transformation to to samples of $\by(\bs_*)$.

\subsection{Random forest model}

An advantage of the Bayesian spatial models is they provide rigorous prediction uncertainties through predictive posterior distributions. These predictive distributions can in turn be absorbed into downstream ecological models that utilize predicted/observed remote sensing metrics such as GEDI RH98, allowing full audits of uncertainties across the modeling workflow.\par

On the other hand, machine learning approaches are attractive as they can model complex relationships between the model covariates and response, often yielding impressive prediction accuracy. This accuracy comes at the expense of theoretical quantification of uncertainty, and therefore perhaps scientific utility \citep{mcroberts2011satellite}, as posing fully probabilistic machine learning models (and subsequently conducting inference) is difficult.\par

We compare the prediction accuracy of the typical and mixture spatial models to a random forest model, an ensemble of randomized decision trees \citep{breiman2001random}. Random forests are popular in remote sensing for both classification and regression tasks \citep{belgiu2016random} and are often used for GEDI data in particular \citep{potapov2021mapping, verhelst2021improving, hoffren2023assessing}. Their popularity is justified by their power to model complex relationships and ease of implementation due to the availability of well-developed and user-friendly software.

We implement the random forests using `R' package `randomForest' \citep{randomForest} with the same Landsat 8 optical values as covariates, using an ensemble of 500 decisions trees, resampling the entire dataset with replacement for every tree. Substantial computational speed-ups were seen when reducing the number of trees and resampling size with only modest detriment to out-of-sample predictions, but we maintained the above parameters to maximize our demonstration of the random forest predictive accuracy. While with the proposed Bayesian models predictive posterior samples of log-RH98 can easily be transformed for predictions of RH98 (or any other transformation), there is no obvious way to transform the random forest predictions. Therefore, we fit two random forest models: one for log-RH98 and another for RH98.

\subsection{Model assessment and comparison}\label{sec:cpo}
Because the objective of the analysis is inference on $\by(\bs_*)$ at unobserved locations $\bs_*$, we use cross-validation and posterior predictive distributions as a means for model assessment and comparison. To evaluate the difference in the predictive distributions of the mixture and typical models, we use conditional predictive ordinate (CPO) scores. Let $s_i$ be the $i$th observation location and $\bs_{-i}$ be the observation locations with the $i$th observation removed. Quantity $\mathrm{CPO}_i$ is the leave-one-out cross-validation predictive density of the $i$th (withheld) observation:
\begin{equation}
\begin{aligned}
\mathrm{CPO}_i = f\left(y(s_i)|\by(\bs_{-i})\right) &= \int f\left(y(s_i)|\bpsi\right)f\left(\bpsi|\by(\bs_{-i})\right)~d\bpsi\\
&= \left(\int \frac{1}{f\left(y(s_i)|\bpsi\right)}f\left(\bpsi|\by(\bs)\right)d\bpsi\right)^{-1},
\end{aligned}
\end{equation}
omitting fixed $\bX(\bs)$ from the notation, and where $\bpsi$ is a vector of the unknown parameters/effects. The final equality above suggests a convenient way to approximate $\mathrm{CPO}_i$ using the posterior samples drawn during the model fit:
\begin{equation}
\mathrm{CPO}_i \approx \left(\frac{1}{M}\sum_{m=1}^M\frac{1}{f\left(y(s_i)|\bpsi_{(m)}\right)} \right)^{-1},
\end{equation}
where $\bpsi_{(m)};~m=1,\ldots,M$ are the $M$ posterior samples of the parameters/effects \citep{held2010posterior}. This avoids the daunting task of refitting the models $n$ times. We use the sum of the log-CPO scores to evaluate predictive fit, which is a proper scoring rule \citep{gneiting2007strictly} that is positively oriented, meaning larger values are evidence that the model better approximates the true predictive distribution. The positive orientation is intuitive, as we desire the posterior predictive density $p\left(y(s_i)|\by(\bs_{-i})\right)$ to be larger at the observed values. Note that the ordering between models of the CPO and total log-CPO scores is invariant to smooth injective transformations of the response. This is important, as different downstream ecological models use different transformations of RH98 as inputs, making the scoring results widely applicable without testing a wide variety of transformations. If we transform posterior predictions of $y(s)$ to $h(s) = g(y(s))$, where $g(\cdot)$ is differentiable and strictly monotone, then using well-known theory on transformation of random variables,
\begin{align}
f_h\left(h(s_i)|\bpsi\right) &=  f\left(y(s_i)|\bpsi\right)\cdot \left| \frac{\partial}{\partial h}g^{-1}\left(h(s_i)\right) \right| = f\left(y(s_i)|\bpsi\right)\cdot c_i \\
\implies  \mathrm{CPO}_{h,i} &\equiv f\left(h(s_i)|\by(\bs_{-i})\right)  = \mathrm{CPO}_{i}\cdot c_i, \\
\text{and}~~~\sum_{i=1}^n \log\left(\mathrm{CPO}_{h,i}\right) &= \sum_{i=1}^n \log\left(\mathrm{CPO}_{i}\right) +  \sum_{i=1}^n \log(c_i).
\end{align}
Constants $c_i;~i\in\{1,\ldots,n\}$ are strictly positive, dependent only on fixed $h(s_i)$, and therefore constant across all compared models, preserving ordering.\par
We also explicitly perform cross-validation according to two different schemes, allowing the random forest to enter the comparison. The first scheme is a random cross-validation, withholding a random sample of 10\% of observations from the model inference and computing predictive distributions (or simply predictions for the random forest) on the withheld test set. For the Bayesian models, we again compare log-densities at the test values and assess the coverage of the 95\% credible intervals, computed using the 2.5\% and 97.5\% posterior quantiles, the coverage for which is also invariant to monotone-increasing transformation. For all models, we use a predictive R-squared metric as a heuristic for the accuracy of the posterior expected values compared to the test values:
\begin{equation}\label{eq:r2}
\tilde{R}^2 = 1 - \frac{\sum_{i=1}^{n_*}\left(\mathrm{E}[y(s_{*,i})] - y(s_{*,i})\right)^2}{\sum_{i=1}^{n_*}\left(y(s_{*,i}) - \bar{y}_*\right)^2 }
\end{equation}
where $s_{*,i};~i\in\{1,\ldots,n_*\}$ are the test locations and $\bar{y}_*$ is the sample mean of the test values.\par
Because GEDI observations are clustered along orbital tracks, the random cross-validation scheme almost guarantees the test locations will be geographically near training locations. This is disadvantageous to the random forest, as random forests do not natively utilize the spatial correlation of the regression residuals. Therefore, we also conduct a by-orbit cross-validation scheme, where subsequently each of the 35 GEDI orbits intersecting the study area are withheld as a test set, leaving the remaining 34 orbits as the training set. This scheme ensures a majority of the test locations will be geographically distant from the training locations. Therefore, each model will primarily rely on the regression with the optical values used as covariates. We use the same metrics and comparisons for the by-orbit scheme as with the previous. \par
We conduct most of the comparison and evaluation on log-RH98, $y(s)$, as the interpretation of most of the results are invariant to transformation and more easily visualized than with extremely-skewed RH98, but explicitly announce when the results are notably different for the original scale, particularly for the $\tilde{R}^2$ values. 

\section{Data Analysis}

Using the data in our study area in Wollemi National Park, Australia, both the typical and mixture spatial models were fit using methods described in Sections \ref{sec:typicalgibbs} and \ref{sec:mixturegibbs}. We drew 3,000 samples for a burn-in followed by an additional 3,000 samples for inference. On a 1.6 GHz machine utilizing 2 cores, the typical model consumed around 20 minutes per 1,000 samples while the mixture model consumed around 90 minutes per 1,000 samples.\par
\begin{figure}[htpb]
	\centering
	\includegraphics[width = \textwidth, keepaspectratio]{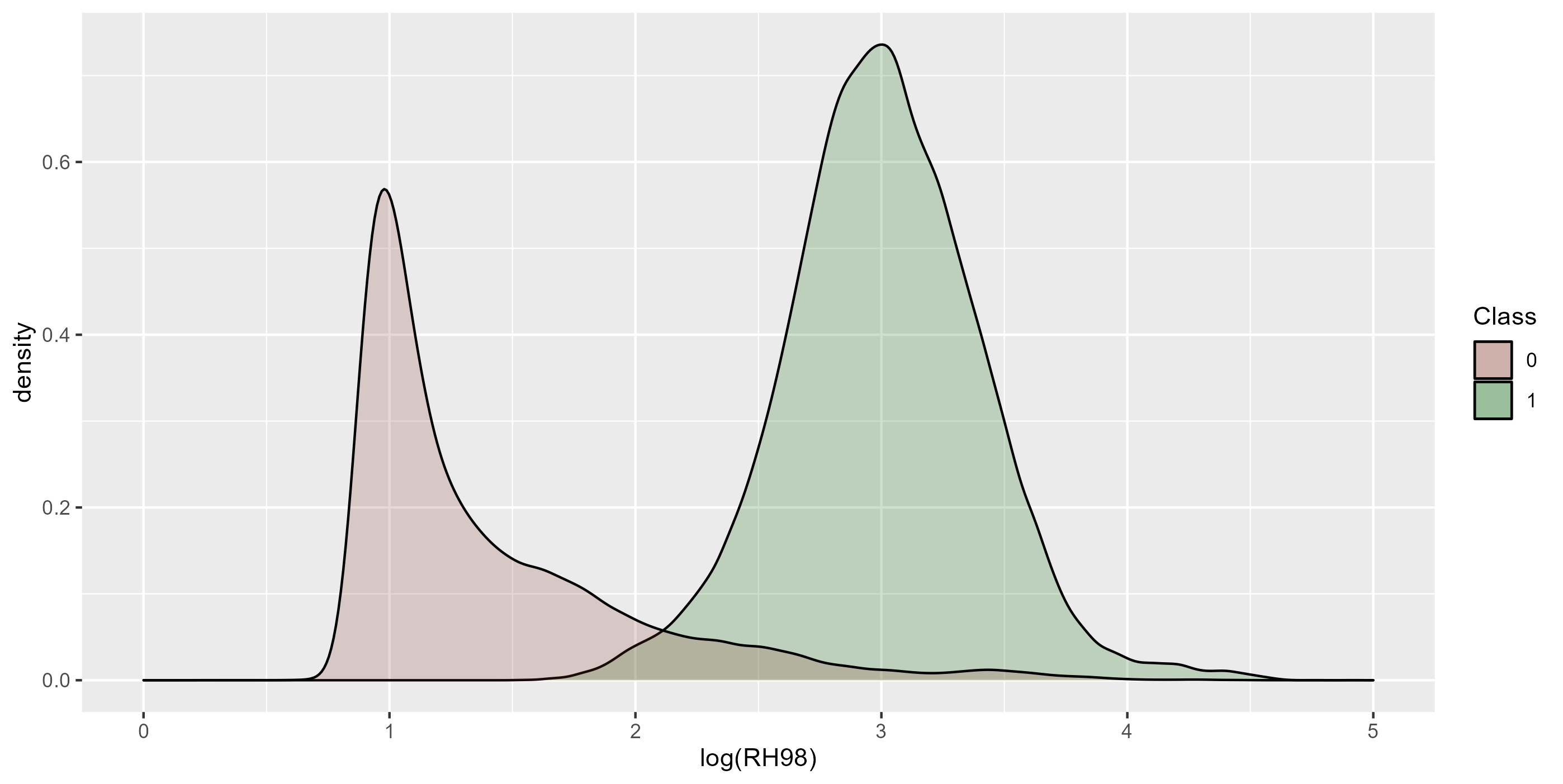}
	\caption{Density plot of the observations, partitioned by their classification, i.e. the posterior mode of $z(s)\in\{0,1\}$. Class 0, the hypothesized non-forest class, has a long right tail reaching into large log-RH98 values.}\label{fig:fpclassdens}
\end{figure}

\begin{figure}[htpb]
	\centering
	\includegraphics[width = \textwidth, keepaspectratio]{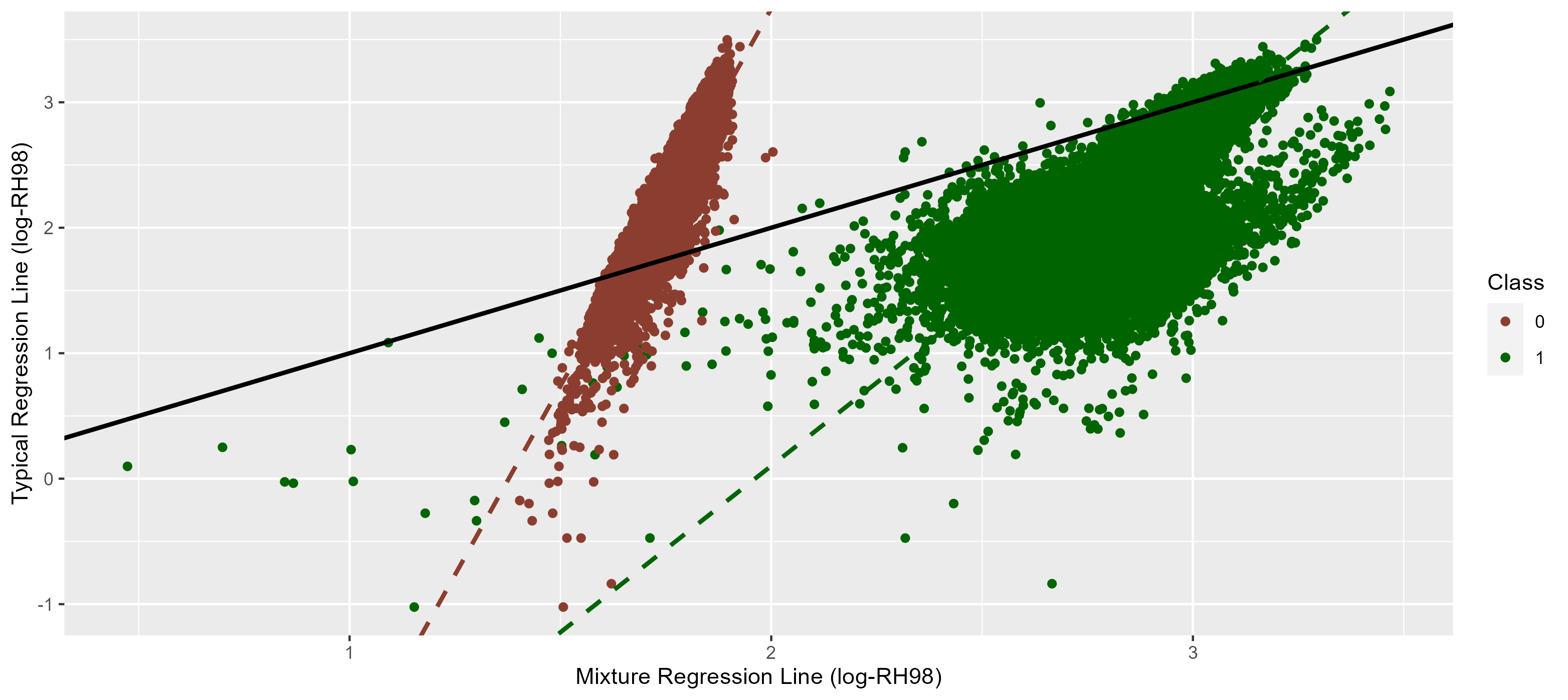}
	\caption{Fitted regression points for the mixture classes $\mathrm{E}[\mu_j + \bX(\bs)\bbeta_j];~j\in\{0,1\}$ versus the fitted regression points for the typical model. Least-squares lines for the mixture classes versus the typical model are drawn through both point clouds, with a one-to-one line in black. The sets of regression points from the two classes are quite different from each other and the one-to-one line, suggesting substantially different relations with the optical intensities across the two classes. Uncertainty for the lines arising from uncertainty of the regression coefficients is negligibly small for the comparison. The more vertical orientation of the class 0 line indicates the relationship between the optical data and log-RH98 is weaker for the non-forest class, which is physically sensible.}\label{fig:reglines}
\end{figure}

 
Table \ref{tab:params} gives posterior values for the both sets of model parameters, excluding the regression coefficients for brevity. For the mixture model, the parameters associated with the two processes, $y_1(s)$ and $y_0(s)$, are quite different. Interestingly, the overall variance of the errors $w_0(s) + \epsilon_0(s)$ is much larger for the hypothesized non-forest class, class 0. This is partially explained by the weaker relationship between the optical data and non-forest process $y_0(s)$ (Figure \ref{fig:reglines}), but not completely. In Figure \ref{fig:fpclassdens}, examining the partition of the observed footprints by their classification (posterior mode of $\bz(\bs)$), the class 0 observations have a long right tail, reaching into large log-RH98 values.\par 
Two possible explanations for these large RH98 values within class 0 are as follows. First, GEDI observations have a geolocation error with a 10 m standard deviation. This leaves the possibility that the observations measured trees of substantial height, but are falsely geolocated at an immediately adjacent location that is clearly non-forested with respect to the optical data, $\bx(s)$. Second, steep slopes result in larger RH98 values even on bare ground. Considering that most of the error magnitude is concentrated in the spatial error $w_0$, which also has a large range ($\sim$4,500 m), the second explanation seems more reasonable. Topographical features such as steep slopes are spatially grouped, whereas large spatial concentrations of the proposed geolocation scenario seem unlikely\par
\begin{table}[htpb]
	\caption{Posterior expected values (standard deviations in parentheses) for the model parameters. The range parameters are given in meters. Variance parameters are converted to standard deviations, e.g. $\sigma = \sqrt{\sigma^2}$. Mean parameters $\mu$ are given with centered covariates and are therefore interpretable as model response means.}\label{tab:params}
	\centering
	\begin{subtable}[b]{\textwidth}
		\centering
		\caption{Mixture model}
		\begin{tabular}{c c c c c c c c c c}
			$\mu_1$ & $\sigma_1$ & $\phi_1$ (m) & $\tau_1$ & $\mu_0$ & $\sigma_0$ &  $\phi_0$ (m) & $\tau_0$ & $\sigma_z$ & $\phi_z$ (m)\\ \hline
			2.90 & 0.31 & 1,614 & 0.27 & 1.79 & 0.62 & 4,468 & 0.22 & 1.51 & 2,018 \\
			(0.01) & (0.004) & (45) & (0.003) & (0.05) & (0.018) & (171) & (0.012) & (0.028) & (72) 
		\end{tabular}
	\end{subtable}\\ \vspace{10pt}
	\begin{subtable}[b]{\textwidth}
		\centering
		\caption{Typical model}
		\begin{tabular}{c c c c}
			$\mu$ & $\sigma$ & $\phi$ (m) & $\tau$ \\ \hline 
			2.49 & 0.44 & 1,833 & 0.46 \\
			(0.03) & (0.005) & (47) & (0.001)
		\end{tabular}
	\end{subtable}
\end{table}

\begin{figure}[htbp]
	\centering
	\includegraphics[width = \textwidth, keepaspectratio]{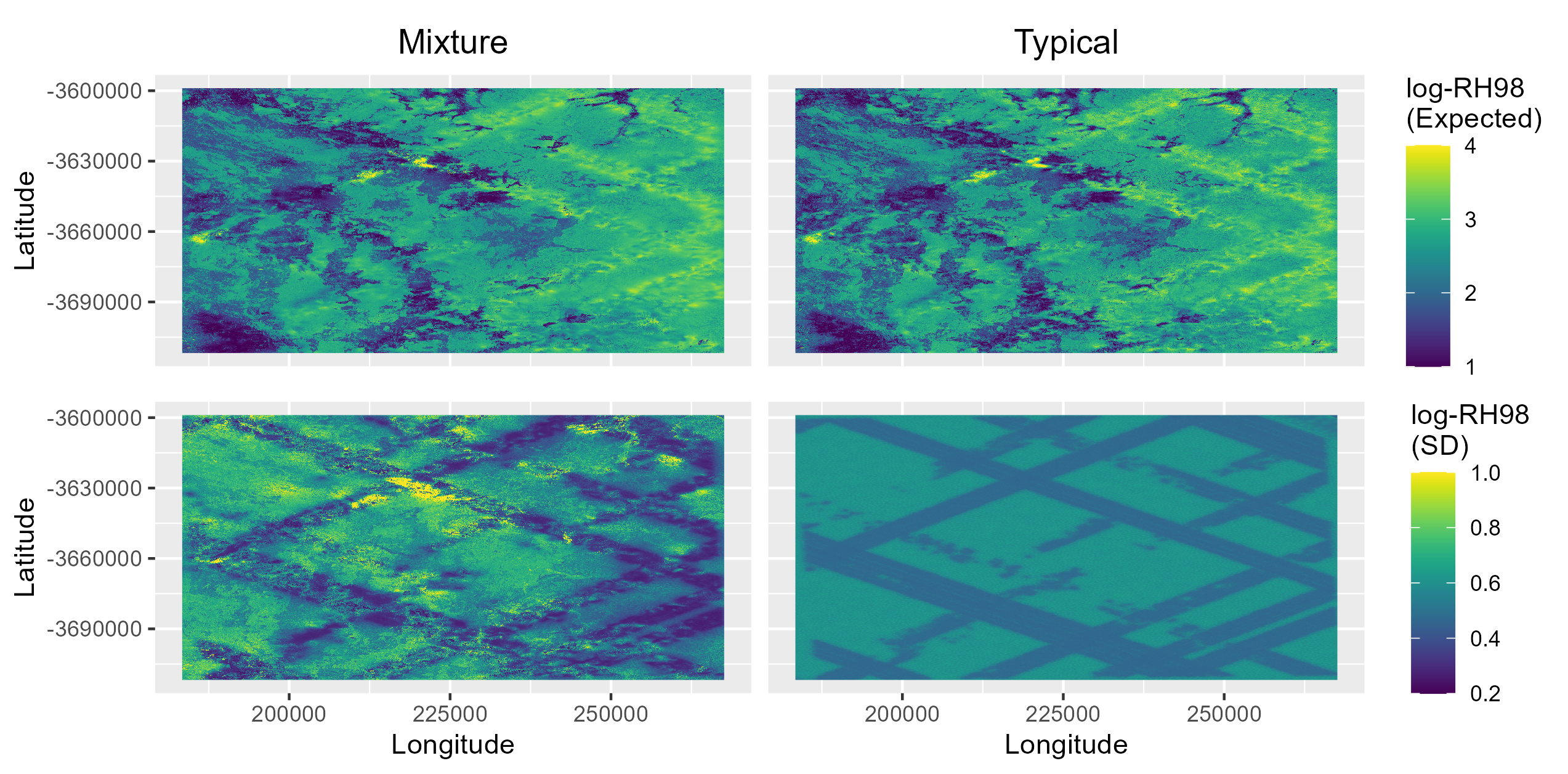}
	\caption{Gridded posterior expected values and standard deviations of $y(s)$ (log-RH98). Mixture model on the left, typical model on the right. Expected values on the top row, standard deviations on the bottom row. The expected values between the mixture and typical model are very similar, but the standard deviations are quite different.}
\end{figure}

\begin{figure}[htbp]
	\centering
	\begin{subfigure}[b]{0.48\textwidth}
		\centering
		\includegraphics[width = \textwidth, height = 1.6in]{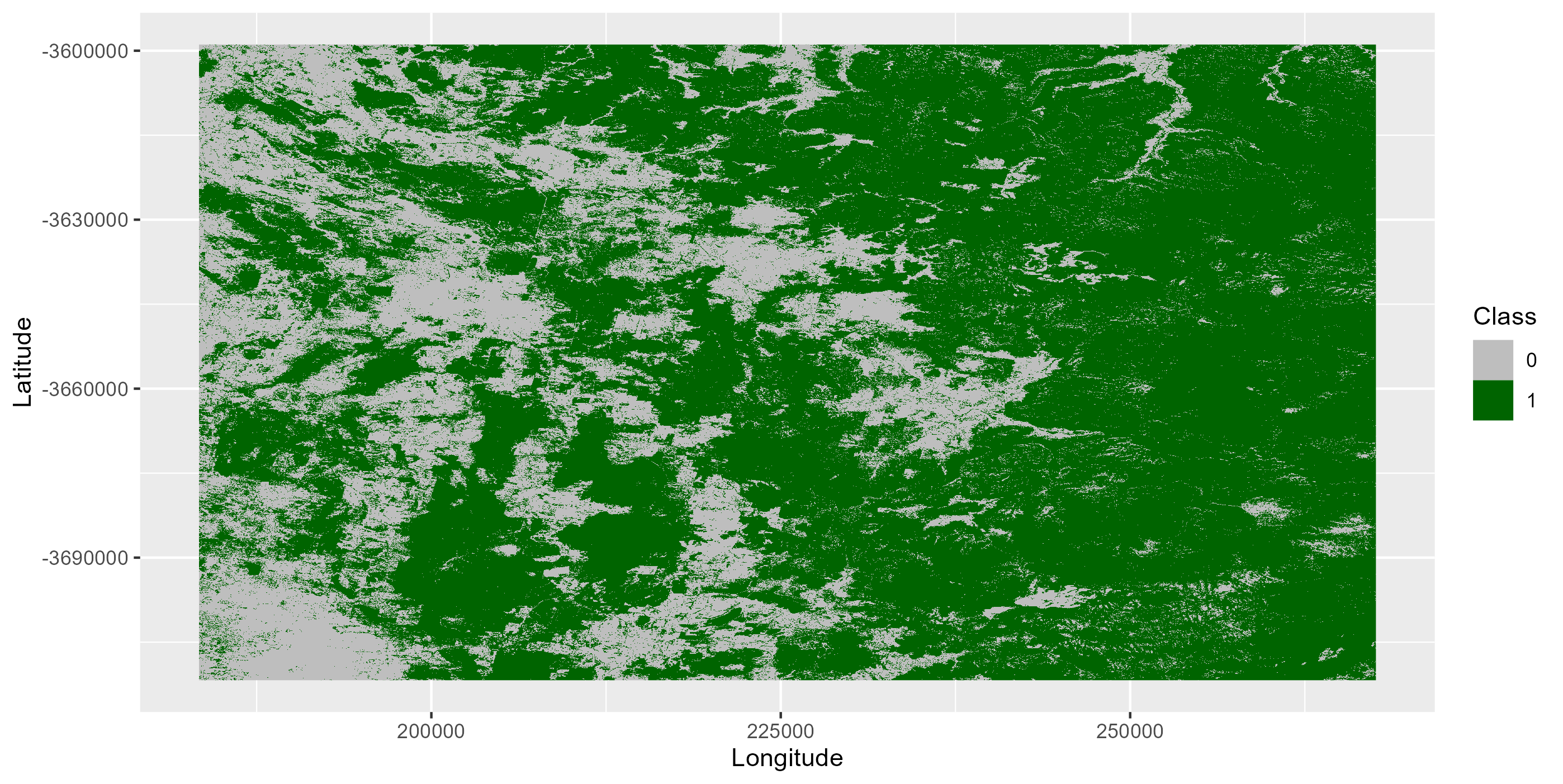}
	\end{subfigure}
	\hfill
	\begin{subfigure}[b]{0.48\textwidth}
	\includegraphics[width = 0.95\textwidth, height = 1.6in]{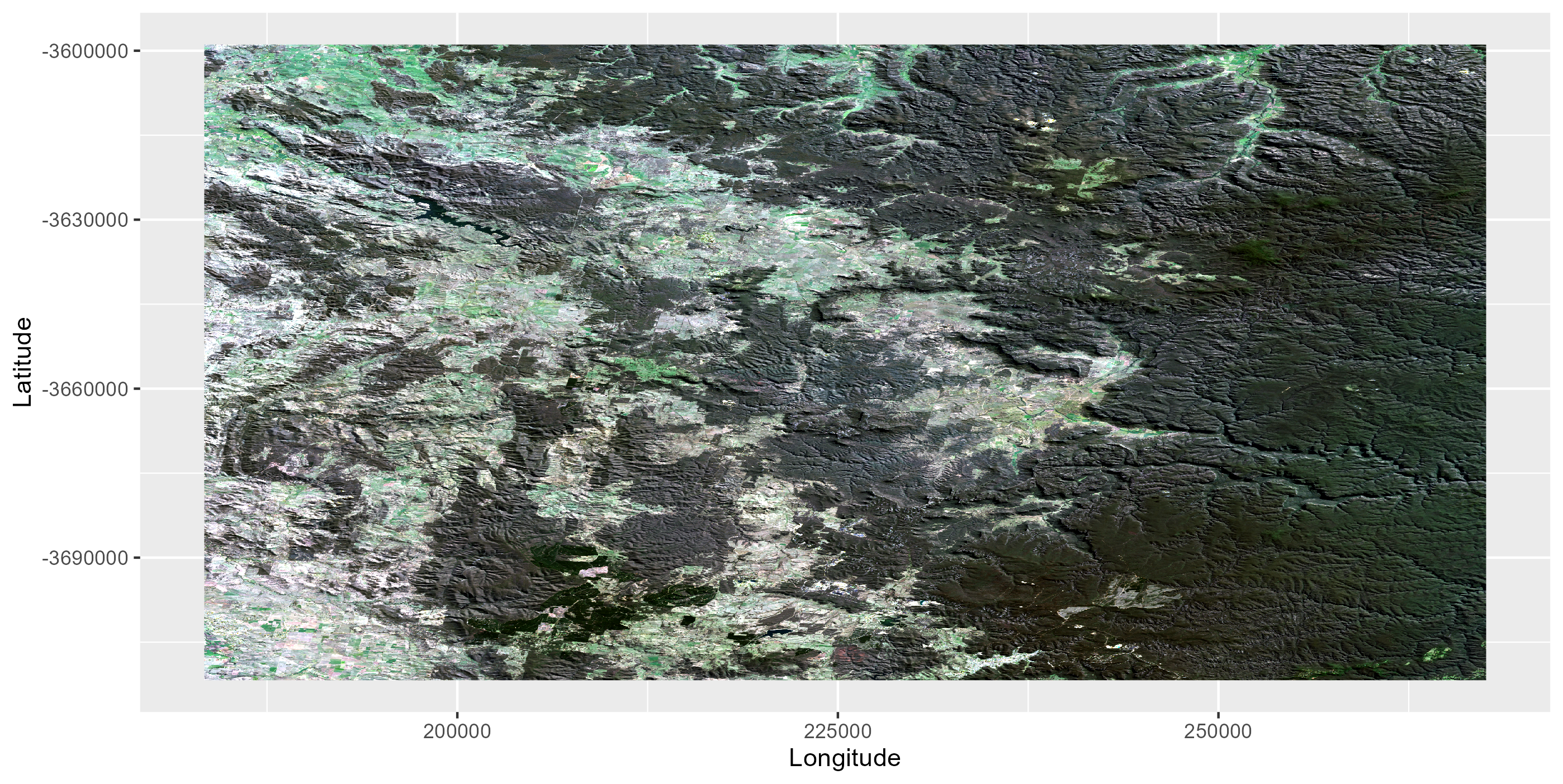}
	\end{subfigure}
	\caption{Comparison of the classification (posterior mode of $z(s)$) compared to an RGB rendering of the Landsat image. The classifications are visibly sensible when compared to the optical imagery.}
\end{figure}

With the posterior samples, posterior predictive samples were computed at the 30 m resolution of the Landsat optical imagery throughout the study domain, constituting over 10 million prediction locations. We examine and compare the predictive inference from both models.\par
\begin{figure}[htbp]
	\centering
	\includegraphics[width = \textwidth, keepaspectratio]{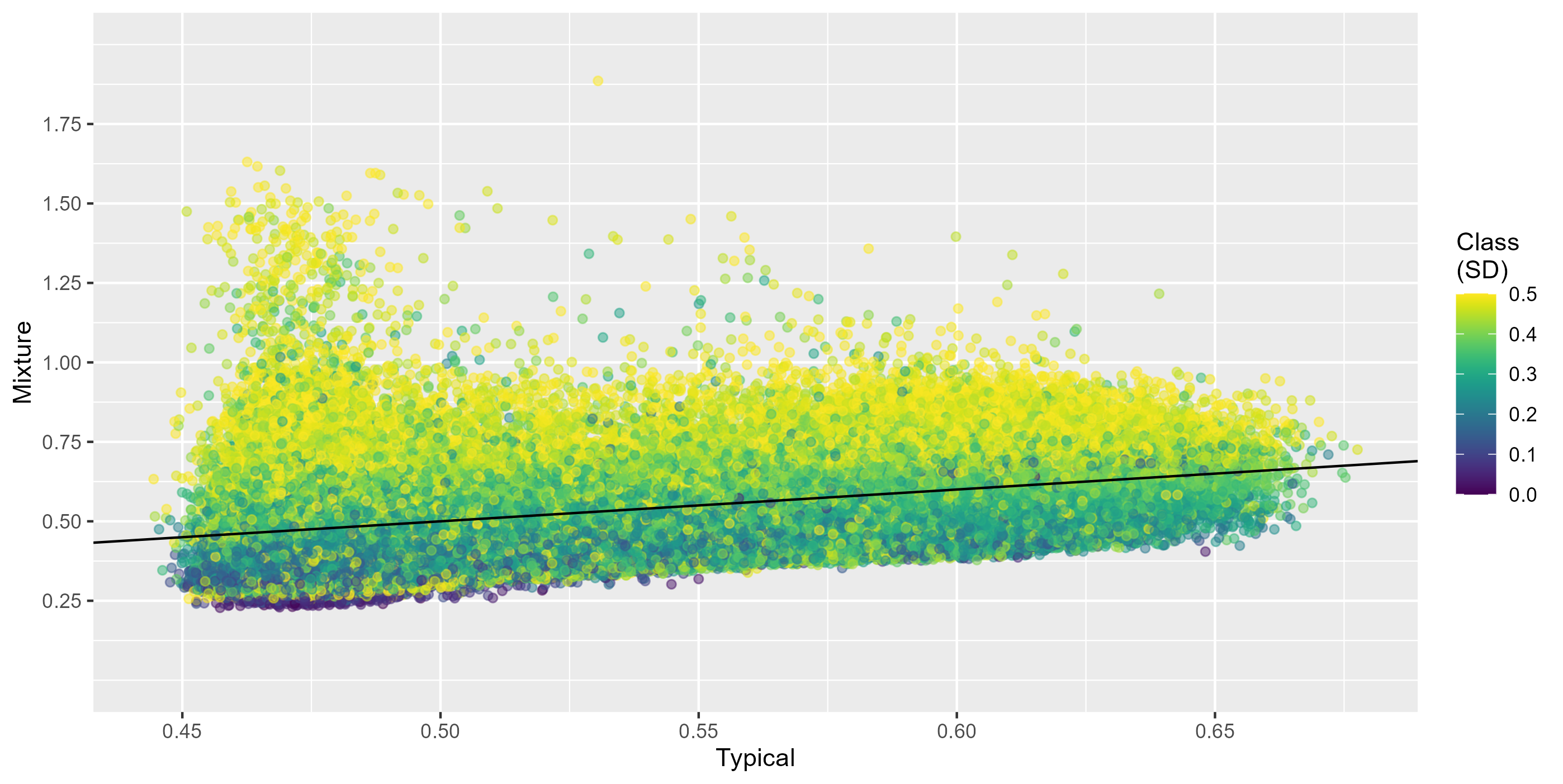}
	\caption{Comparison of the posterior predictive standard deviations for the mixture and typical models, with a one-to-one line in black, using a random sample of 100,000 locations out of the $\sim$10 million prediction locations. Points are colored by the posterior standard deviation of class $z(s)$, showing that on average the mixture model has lower standard errors when class-certainty is high, and higher standard errors when class-certainty is low. }\label{fig:sd_vs_classsd}
\end{figure}

 
A categorical difference between the two models is the mixture model provides inference on the classification of the prediction locations. Comparing the classified grid to the Landsat imagery, the locations with a posterior mode of $z(s) = 1$ accord well with the obviously forested areas. The posterior expected values for log-RH98 between the typical and mixture models are similar, with a mean squared difference of 0.016, which is around 4\% of the sample variance of the expected values from either model. These slight differences in the expected values are exacerbated to a more notable 10\% when posterior samples are transformed to the original RH98 scale. However, the mixture model's posterior predictive distributions are thoroughly distinguished in every other respect. The typical model's posterior standard deviations are entirely driven by the prediction location's proximity to the GEDI orbits, whereas those of the mixture model exhibit substantial variation beyond proximity to the orbits. Indeed, a primary driver of the standard deviations for the mixture model is the certainty of the classification (Figure \ref{fig:sd_vs_classsd}). For locations where $z(s)$ is confidently estimated, the mixture model overall gives lower standard deviations on $y(s)$. For locations where $z(s)$ is uncertain, the opposite is true.\par
\begin{figure}[htpb]
	\centering
	\includegraphics[width = \textwidth, keepaspectratio]{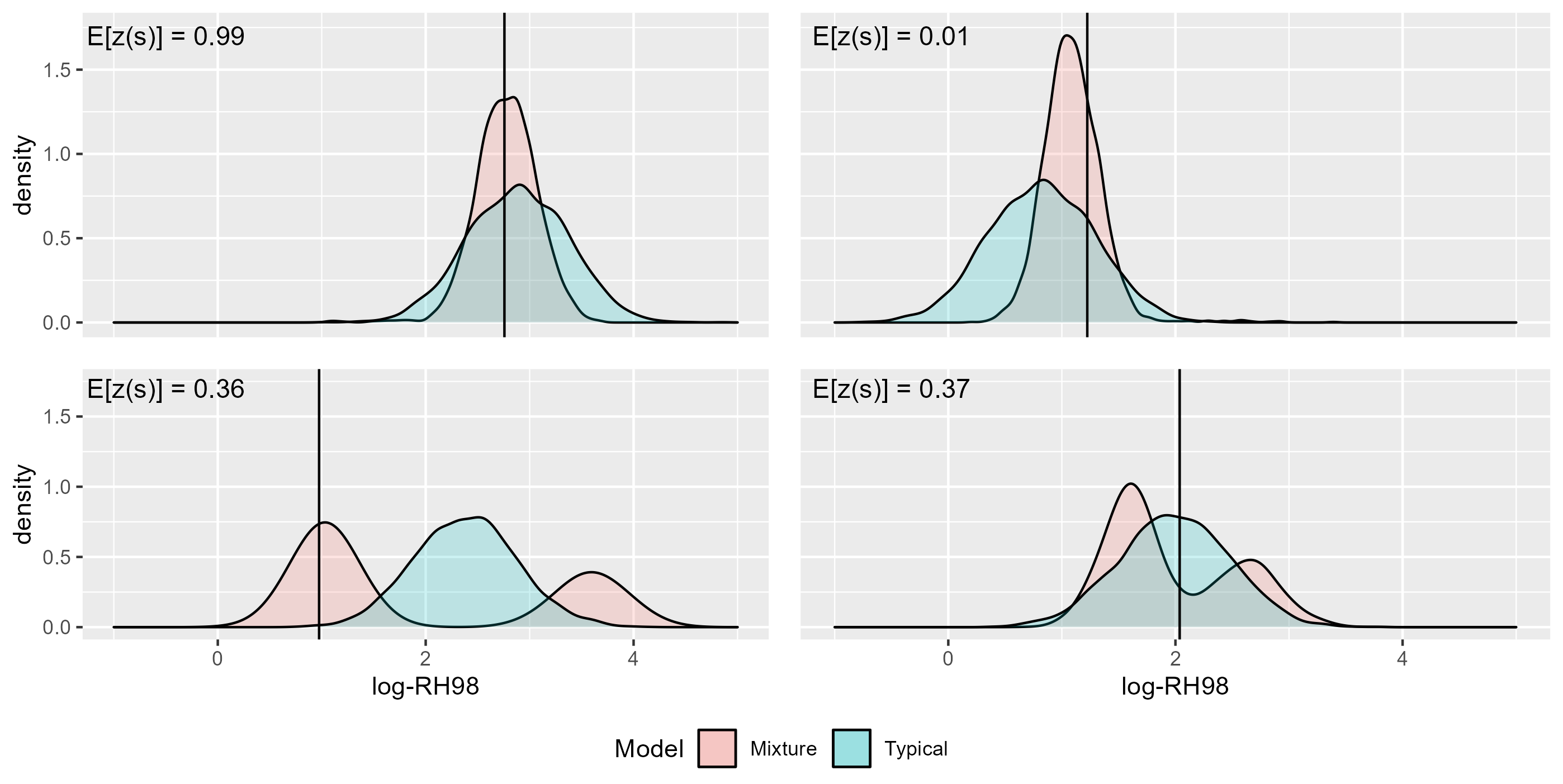}
	\caption{Comparison of the posterior predictive distributions for four example locations in the cross-validation, with vertical black lines on the true value. The top row gives two cases where the class certainty is high, where the mixture posterior is unimodal but narrower than the typical posterior. The bottom row gives two cases where class membership is uncertain and the mixture model has a bimodal posterior. The bottom left shows an instance where the bifurcation gives much higher posterior density at the true value than the typical model. The bottom right shows the rarer instance where the bifurcation gives a lower posterior density. }
	\label{fig:cvpostdens}
\end{figure}

Computing the total log-CPO scores of Section \ref{sec:cpo}, the mixture model yielded a value of -33,299 while the typical model yielded -63,222, evidence that the mixture model produces superior predictive distributions. To further explore the differences in the predictive densities and values, we perform explicit cross-validation, introducing the random forest in the comparison. We first demonstrate the random cross-validation, withholding a random 10\% of the observations, fitting the models to the remainder and computing predictions for the withheld test observations. Again, the posterior expected values for log-RH98 were similar between the mixture and typical models, with an $\tilde{R}^2$, (\ref{eq:r2}), between the expected and withheld values of $0.71$ for the mixture and $0.70$ for the typical model. For this cross-validation scheme, the accuracy of the random forest was substantially lower than the spatial models, with an $\tilde{R}^2$ of $0.60$. When transforming posterior samples (and the test values) to the original RH98 scale, and fitting a new random forest model for untransformed RH98, the differences are expanded to a $\tilde{R}^2$ of $0.60$ for the mixture model, $0.53$ for the typical model, and $0.41$ for the random forest. The mixture model achieves much higher predictive densities than the typical model, with a total log-density of -3,302 compared to a total log-density of -6,617 for the typical model. The bifurcation into two mixed processes is on average beneficial. Figure \ref{fig:cvpostdens} illustrates this with predictive posteriors at example test locations. For locations where the classification $z(s)$ is certain, the mixture model yields near-unimodal predictive densities with less dispersion than the typical counterpart. For locations where the classification is uncertain, the mixture model often ``hedges its bets'' with a distinctly bimodal distribution that on average places higher density on the true value when compared to the typical model's single unimodal distribution. For some locations, the class models $y_1(s),~y_0(s)$ have similar expected values, in which case a unimodal distribution results, even when the class probability is near $0.5$. The coverage rate of the 95\% equal-tail credible intervals was near nominal for both spatial models, with 95.3\% for the mixture model and 93.7\% for the typical model.\par

We next present the by-orbit cross-validation described in Section \ref{sec:cpo}, where the majority of test locations are geographically distant from the training locations. Here, the random forest exhibits substantially better predictive accuracy than either spatial models, with a $\tilde{R}^2$ of $0.55$, $0.52$ and $0.60$ for the mixture, typical, and random forest model respectively for log-RH98, and a $\tilde{R}^2$ of $0.32$, $0.28$ and $0.39$ respectively for RH98. The relative performance between the mixture and typical models remains similar, with a total log-density of -66,154 and -95,716 for the mixture and typical model respectively. Again, the coverage rate of the 95\% credible intervals was near nominal for both spatial models at 94.7\% and 93.2\% respectively.

\section{Discussion}

The spatial mixture model accounts for categorically different properties of heterogeneously distributed classes across a study domain. For our data, this lead to higher predictive densities at the true values when compared to a single spatial model. The higher predictive densities are achieved through tighter, class-specific, posterior distributions when the class certainty is high, and a bimodal posterior distributions that acknowledges two discrete possibilities when the class certainty is low. However, considering scalar predictions made through posterior expected values, there are only slight differences between the mixture and single spatial model, even upon transforming from the log-scale to the original RH98 scale. In the cross-validation studies, the sample standard error between the expected RH98 and test values was improved by $\sim$0.6 m using the mixture model, which may or may not be deemed physically significant. If only scalar predictive maps of RH98 are desired, an argument could be made to avoid the additional computational expenditure of the mixture model. Indeed, in this case, an argument could be made for avoiding model-based geostatistics altogether in favor of algorithmic, machine-learning approaches such as the random forest. The random forest exhibited substantially better predictive accuracy when making predictions geographically distant from training locations. This is due to the random forest's ability to better capture complex relations between the passive optical covariates and (log)-RH98 than the linear regressions in the spatial models. Even the spatial models advantage in accuracy for geographically near predictions could likely be removed by subsequently performing kriging on the residuals of the random forest. But if accounting for and auditing uncertainties is a priority, the rich posteriors of the spatial mixture model paired with its reasonable accuracy seem advantageous.\par
Potentially one of the greatest advantages of the mixture model is the separation of observations and predictions into classes, which may drastically alter the interpretation of the predicted scalar values and their absorption in downstream modeling. We hypothesized the two classes to be forest and non-forest. While this hypothesis seems reasonable, given the orientation of the classes and the comparison to the optical image, there was no labeling in the training data to enforce this, nor ground-truth data for testing. Rather, the classification is `unsupervised', separating data into classes that are the most internally cohesive and externally separated. Further investigation is required to deduce the accordance of the classes to ground-truth definitions of forest, though these definitions are often nebulous and somewhat arbitrary (is a plot with a single 2 m sapling considered `forested'?).\par
Further, we assumed a fixed number of classes, two, which was well-motivated from our data and prior knowledge of the study area: Wollemi National Park is a reserved area, composed primarily of forest and exposed heath/grassland. This assumption could be relaxed to an arbitrary number of $K$ classes using the techniques in \cite{neelon2014multivariate, vanhatalo2021spatiotemporal}, where categorical probabilities are given by the softmax transformation of $K$ latent spatial processes. In this case, greater care must be taken to ensure identifiability of the class labels. Additionally, there is no guarantee that all ecologically meaningful separations would manifest in the unsupervised classification of the mixture model: the distinct classes would need to produce significantly different distributions of observed GEDI metrics. \par
We assumed that upon a log-transformation, the observations followed a mixture of linear spatial models with normally distributed random effects. While this seemed reasonable upon data exploration (Figure \ref{fig:datadescrip}), model inference revealed that the non-forest class had a large right tail in the distribution (Figure \ref{fig:fpclassdens}). The assumption could be relaxed with a mixture of generalized linear spatial models \citep[Section 5.2]{banerjee2003hierarchical}. For instance, a mixture of two Gamma spatial models with different dispersion parameters could account for varying skews in the class distributions. This would preclude closed-form conditional samples of the latent effects, requiring either more Laplace approximations or careful schemes balancing computational feasibility and exact Bayesian inference. Evidence for a more flexible model, such as a mixture of Gamma models, could be assessed through a Bayes factor, or through superior predictive performance, as indicated by log-CPO scores.\par

\nocite{*} 

\appendix

\section{Gibbs samplers}\label{app:gibbs}

Here we give exact details on the Gibbs sampler for the typical and mixture model. First presenting our notation, let $\bs= [s_1 \cdots s_n]^T$ be the vector of $n$ observation locations, $\by(\bs)$ be the vector of observed log-RH98 and $\bX(\bs)$ be the $n\times p$ matrix of optical intensities. 

\subsection{Typical spatial model}

The Gibbs sampler for the typical spatial model iterates as follows:
\begin{enumerate}
\item Sample spatial effects $\bw|\by(\bs), \bX(\bs), \mu, \bbeta,\theta, \tau^2$, which is multivariate normal with mean and variance
\begin{align}
\mathrm{E}[\bw| \ldots] &=(\bQ(\theta) + \frac{1}{\tau^2}\bA(\bs)^T\bA(\bs))^{-1}\bA(\bs)^T\frac{1}{\tau^2}(\by(\bs) - \mu\boldsymbol{1} - \bX(\bs)\bbeta),\label{eq:wmu}\\ 
\mathrm{Var}[\bw| \ldots] &= (\bQ(\theta) + \frac{1}{\tau^2}\bA(\bs)^T\bA(\bs))^{-1}.
\end{align}
\item Sample jointly mean parameters $\mu,\bbeta| \by(\bs), \bX(\bs), \theta, \tau^2$ marginalizing over $\bw$, which is multivariate normal with mean and variance 
\begin{align}
\mathrm{E}\begin{bmatrix} \mu \\ \bbeta \end{bmatrix} &= \left(\tilde{\bX}(\bs)^T\bSigma(\bs;\theta)^{-1}\tilde{\bX}(\bs)\right)^{-1}\tilde{\bX}(\bs)^T\bSigma(\bs;\theta)^{-1}\by(\bs), \\
\mathrm{Var}\begin{bmatrix} \mu \\ \bbeta \end{bmatrix} &= \left(\tilde{\bX}(\bs)^T\bSigma^{-1}\tilde{\bX}(\bs)\right)^{-1},
\end{align}
where $\bSigma(\bs;\theta) = \bA(\bs)\bQ(\theta)^{-1}\bA(\bs)^T + \tau^2\bI$ and $\tilde{\bX}(\bs) = [\bone,~\bX(\bs)]$. Using the Woodbury matrix identity,
\begin{align}
\bSigma^{-1} = \frac{1}{\tau^2}\bI - \frac{1}{\tau^4}\bA(\bs)^T\left(\bQ(\theta) + \frac{1}{\tau^2}\bA(\bs)^T\bA(\bs)\right)^{-1}\bA(\bs).
\end{align}
\item Sample the noise variance $\tau^2|\by(\bs), \bX(\bs), \mu, \bbeta,\bw$, which is inverse-gamma with shape $n/2$ and rate
\begin{equation}\label{eq:igvariance}
\frac{1}{2}\left(\by(\bs) - \mu\boldsymbol{1} - \bX(\bs)\bbeta - \bA(\bs)\bw\right)^T\left(\by(\bs) - \mu\boldsymbol{1} - \bX(\bs)\bbeta - \bA(\bs)\bw\right).
\end{equation}
\item Sample jointly the Mat\'ern parameters $\theta = \{\sigma^2, \phi\}$ using a Metropolis-Hastings step on target density
\begin{equation}\label{eq:thetapost}
p(\theta|\bw) \propto |\bQ(\theta)|^{1/2}\exp\left(-\frac{1}{2}\bw^T\bQ(\theta)\bw\right).
\end{equation} 
\end{enumerate}

\subsection{Spatial mixture model}\label{app:mixturegibbs}

Let $\bs_j = \{s_i~:~z(s_i) = j\}$ for $j\in\{0,1\}$ be the sub-vector of observations corresponding to either class. The Gibbs sampler for the spatial mixture model iterates as follows.
\begin{enumerate}
\item Sample classifications $\bz(s_i)|\ldots;~i\in\{1,\ldots,n\}$, which are Bernoulli distributed with conditional probabilities
\begin{equation}
\pi^*(s_i) = \frac{\pi(s_i)f\left(y(s_i)| z(s_i) =1\right)}{\pi(s_i)f\left(y(s_i)| z(s_i) =1\right) + (1-\pi(s_i))f\left(y(s_i)| z(s_i) =0\right)},
\end{equation}
where $f\left(\cdot|z(s_i) = j\right)$ is a normal density with mean $\mu_j + \bx(s_i)^T\bbeta_j + \bA(s_i)\bw_j$ and variance $\tau_j^2$ for $j\in\{0,1\}$, and $\pi(s_i)$ is equation (\ref{eq:pprocess}) evaluated at current parameter and effect values.
\item Sample spatial effects $\bw_j|\by(\bs_j), \bX(\bs_j), \mu_j, \bbeta_j,\theta_j, \tau^2_j$ for $j\in\{0,1\}$, which are multivariate normal with mean and variance
\begin{align}
\mathrm{E}[\bw_j| \ldots] &=(\bQ(\theta_j) + \frac{1}{\tau_j^2}\bA(\bs_j)^T\bA(\bs_j))^{-1}\bA(\bs_j)^T\frac{1}{\tau_j^2}(\by(\bs_j) - \mu_j\boldsymbol{1} - \bX(\bs_j)\bbeta_j),\\ 
\mathrm{Var}[\bw_j| \ldots] &= (\bQ(\theta_j) + \frac{1}{\tau_j^2}\bA(\bs_j)^T\bA(\bs_j))^{-1}.
\end{align}
\item Sample jointly mean parameters $\mu_j,\bbeta_j| \by(\bs_j), \bX(\bs_j), \theta_j, \tau_j^2$ marginalizing over $\bw_j$ for $j\in\{0,1\}$, which are multivariate normal with mean and variance 
\begin{align}
\mathrm{E}\begin{bmatrix} \mu_j \\ \bbeta_j \end{bmatrix} &= \left(\tilde{\bX}(\bs_j)^T\bSigma(\bs_j;\theta_j)^{-1}\tilde{\bX}(\bs_j)\right)^{-1}\tilde{\bX}(\bs_j)^T\bSigma(\bs_j;\theta_j)^{-1}\by(\bs_j), \\
\mathrm{Var}\begin{bmatrix} \mu_j \\ \bbeta_j \end{bmatrix} &= \left(\tilde{\bX}(\bs_j)^T\bSigma(\bs_j;\theta_j)^{-1}\tilde{\bX}(\bs_j)\right)^{-1},
\end{align}
where $\bSigma(\bs_j;\theta_j) = \bA(\bs_j)\bQ(\theta_j)^{-1}\bA(\bs_j)^T + \tau_j^2\bI$ and $\tilde{\bX}(\bs_j) = [\bone,~\bX(\bs_j)]$. Using the Woodbury matrix identity,
\begin{align}
\bSigma(\bs_j;\theta_j)^{-1} = \frac{1}{\tau_j^2}\bI - \frac{1}{\tau_j^4}\bA(\bs_j)^T\left(\bQ(\theta_j) + \frac{1}{\tau_j^2}\bA(\bs_j)^T\bA(\bs_j)\right)^{-1}\bA(\bs_j).
\end{align}
\item Sample the noise variance $\tau_j^2|\by(\bs_j), \bX(\bs_j), \mu_j, \bbeta_j,\bw_j$ for $j\in\{0,1\}$, which are inverse-gamma with shape $n_j/2$ (where $n_j$ is the length of $\bs_j$) and rate
\begin{equation}\label{eq:igvariance}
\frac{1}{2}\left(\by(\bs_j) - \mu_j\boldsymbol{1} - \bX(\bs_j)\bbeta_j - \bA(\bs_j)\bw_j\right)^T\left(\by(\bs_j) - \mu_j\boldsymbol{1} - \bX(\bs_j)\bbeta_j - \bA(\bs_j)\bw_j\right).
\end{equation}
\item Sample jointly $\mu_p,\bbeta_p, \bw_p| \bz(s), \theta_z$ using a Laplace approximation. Let $\bb =  [\mu_p,\bbeta_p, \bw_p]^T$ and define design matrix $\tilde{\bA}=[\bone, \bX(\bs),\bA(\bs)]$ and the prior precision $\tilde{\bQ}(\theta_p) = \mathrm{blockdiag}\{\bzero_{p+1}, \bQ(\theta_p)\}$, where $\bzero_{p+1}$ is a $p+1\times p+1$ matrix of zeros, representing the prior precision of the scalar intercept $\mu_z$ and $p$ regression coefficients $\bbeta_z$. A Newton-Raphson routine is used to find posterior mode $\hat{\bb}$. Letting  $\bb_{(0)}$ be some initial value, we iterate
\begin{align}
\bb_{(j+1)} &= \bb_{(j)} - \bH\left(f(\bb_{(j)}|\bz(\bs),\theta)\right)^{-1}\frac{\partial f}{\partial\bb}(\bb_{(j)}|\bz(\bs),\theta)\\
&= \bb_{(j)} + \left(\tilde{\bQ}(\theta_p) + \tilde{\bA}^T\bD(\bb_{(j)})\tilde{\bA}\right)^{-1}\left(\tilde{\bA}^T\left(\bz(\bs) - \bp(\bs|\bb_{(j)})\right) - \tilde{\bQ}(\theta_p)\bb_{(j)}\right),
\end{align}
where $\bD(\bb_{(j)}) = \mathrm{diag}\{\pi(s_i|\bb_{(j)})(1 - \pi(s_i|\bb_{(j)})\}$, until a convergence criterion is met. Then the Laplace approximation is
\begin{equation}
\bb|\bz(s),\theta_z \sim \mathrm{MVN}\left(\hat{\bb},~~\left(\tilde{\bQ}(\theta_z) + \tilde{\bA}^T\bD(\hat{\bb})\tilde{\bA}\right)^{-1}\right).
\end{equation}
\item Sample jointly the Mat\'ern parameters $\theta_j = \{\sigma_j^2, \phi_j\}$ for $j\in\{0,1,z\}$, using Metropolis-Hastings steps on the target densities
\begin{equation}\label{eq:thetapost}
p(\theta_j|\bw_j) \propto |\bQ(\theta_j)|^{1/2}\exp\left(-\frac{1}{2}\bw_j^T\bQ(\theta_j)\bw_j\right).
\end{equation} 
\end{enumerate}

\end{document}